\documentclass[sn-mathphys]{sn-jnl}% Math and Physical Sciences Reference Style
\jyear{2021}%

\theoremstyle{thmstyleone}%
\theoremstyle{thmstyletwo}%

\theoremstyle{thmstylethree}%
\usepackage[T1]{fontenc}

\begin{document}

\title[Nutzungsverhalten und Funktionsanforderungen digitaler Trainingsanwendungen]{Nutzungsverhalten und Funktionsanforderungen digitaler Trainingsanwendungen während der Pandemie}


\author[1]{\fnm{Freya} \sur{Pfau}}\email{freya\_schulte@web.de}
\author[2]{\fnm{Johannes} \sur{Pfau}}\email{jopfau@ucsc.edu}
\author[3]{\fnm{Bastian} \sur{Dänekas}}\email{daenekba@uni-bremen.de}
\author[3]{\fnm{Robert} \sur{Porzel}}\email{porzel@uni-bremen.de}
\author[3]{\fnm{Rainer} \sur{Malaka}}\email{malaka@uni-bremen.de}
\author[1]{\fnm{Melanie} \sur{Krüger}}\email{melanie.krueger@sportwiss.uni-hannover.de}

% \author{Anonyme Autoren}

\affil[1]{\orgdiv{Institut für Sportwissenschaft}, \orgname{Universität Hannover}, \orgaddress{\street{Am Moritzwinkel 6}, \city{Hannover}, \postcode{30167}, \country{Deutschland}}}
\affil[2]{\orgname{University of California, Santa Cruz}, \orgaddress{\street{1156 High St}, \city{Santa Cruz}, \postcode{95064}, \country{USA}}}
\affil[3]{\orgname{Universität Bremen}, \orgaddress{\street{Bibliothekstraße 1}, \city{Bremen}, \postcode{28359}, \country{Deutschland}}}

% \affil[2]{\orgdiv{Department}, \orgname{Organization}, \orgaddress{\street{Street}, \city{City}, \postcode{10587}, \state{State}, \country{Country}}}

% \affil[3]{\orgdiv{Department}, \orgname{Organization}, \orgaddress{\street{Street}, \city{City}, \postcode{610101}, \state{State}, \country{Country}}}


\abstract{Durch unter anderem Kontaktverbote, Schließung von Fitnesseinrichtungen und Quarantänemaßnahmen führte die SARS-CoV-2 Pandemie zu einem erheblichen Rückgang sportlicher Aktivitäten. Die erste Lockerung oder Aufhebung dieser Einschränkungen führte in Deutschland zu einem weitestgehend vollständigen Zurückkehren der Normalität des Bewegungs- und Trainingsverhaltens, die Langzeitwirkung der wiederkehrenden Maßnahmen (speziell für Deutschland des ``Lockdowns'', ``Lockdown lights'' und der ``Corona Notbremse'') wurden bislang allerdings noch nicht ausreichend untersucht. Über eine Umfrage von (n=108) deutscher Sportler*innen wurde in dieser Arbeit ein signifikanter Rückgang der sportlichen Aktivität festgestellt, selbst in Zwischenphasen ohne pandemische Einschränkungen. Um die Chancen digitaler Trainingsanwendungen diesem entgegenzuwirken festzustellen, wurde deshalb die Nutzung von unter anderem Apps, Trackern, Videoangeboten und Exergames erfasst und die wichtigsten Features, sowie fehlende oder sogar notwendige Funktionen erhoben, die individuellen Sport oder Training auch über Zeiträume ohne nutzbare Einrichtungen oder Kontakt ermöglichen würden. Tatsächlich nahm die Nutzung von Smartwatches, Videoangeboten und Videokonferenzen zur sportlichen Unterstützung im Vergleich zu vor der Pandemie signifikant zu und insbesondere Videoangebote und –konferenzen trugen dazu bei, die Trainingshäufigkeit zu erhöhen. Datengetriebenes oder persönliches Feedback, Motivation und Kollaboration stellen hierbei die Funktionen dar, die Nutzer*innen digitaler Trainingsanwendungen als am wichtigsten bewerten oder sogar voraussetzen und dem Rückgang der sozialen Komponente des Trainings entgegenwirken können.}
% 199 wörter (150-250 maximal)




\maketitle
% German Journal of Exercise and Sport Research
% Sportwissenschaft

% 5000 words (~24 pages)
% 50 References (APA Style)
% 5 Tables and/or Figures

% 150-250w Abstract
% 4-6 Keywords
\section{Einleitung}
% Related Work mit in Enleitung

Durch das globale Auftreten der SARS-CoV-2 Pandemie \cite{Wu2020} und der darauffolgenden Einschränkungsmaßnahmen hinsichtlich Kontaktverbot, Schließung von Einrichtung und Quarantänen \cite{Imoehl2021} ist sowohl professioneller als auch Freizeitsport und persönliches Training signifikant zurückgegangen.
In Deutschland nahm bei 31\% der deutschen Erwachsenen die sportliche Freizeitaktivität ab, bei 6\% zu und 27\% behielten ihr persönliches Niveau bei \cite{Mutz2021, Mutz2021b}. Koopmann et al. bestätigen diese Entwicklung mit einem sportlichen Rückgang von 38,3\% unter den Teilnehmenden \cite{Koopmann2021}, laut Sondhof et al. bezeichnen sogar 40\%  deutscher Studierender ihre körperliche Verfassung als schlechter als vor der Pandemie \cite{Sondhof2020}. Auch unter der älteren Bevölkerung gab etwa ein Drittel an, seit der Pandemie weniger Sport zu treiben \cite{Nowossadeck2021}. Darüber hinaus litten deutsche Sportvereine  unter einem drastischen Rückgang von Mitgliedschaften ab 2020 \cite{Thieme2021} und laut dem Arbeitgeberverband deutscher Fitness- und Gesundheits-Anlagen sind die Fitnessmitgliedschaften und Umsatzentwicklung der Branche drastisch zurückgegangen \cite{DSSV2020}.
Rubach et al. verzeichneten eine leichte Abnahme der Bewegungsintensität aller Altersgruppen in Deutschland anhand des Rückgangs wöchentlicher Häufigkeiten leichter, mittlerer und starker Bewegungen \cite{Rubach2021}.
Der Hersteller der Fitnesstracker \textit{Fitbit} veröffentlichte einen 7-prozentigen Rückgang der Schrittzahlen unter allen deutschen Benutzer*innen in den ersten 3 Monaten der Pandemie, welche allerdings im europäischen Bereich noch am besten abschnitten \cite{Fitbit2020}.
In Ländern mit ähnlichen Einschränkungen (Kanada, Südkorea, Italien) sind sportliche Aktivitäten unter Kindern und Jugendlichen zurückgegangen und Bildschirmzeit (Fernseher, Computer, Smartphone) entsprechend deutlich gestiegen \cite{Guan2020, Pietrobelli2020}. Unter australischen Studierenden wurde ebenso ein Rückgang der körperlichen Aktivitäten um etwa $30\%$ seit Beginn der Pandemie festgestellt \cite{Gallo2020}.
Auch wenn Sportler*innen nach momentanem medizinischen Stand nicht zu den Risikogruppen für einen schweren COVID-19-Verlauf gehören, kann auch bei gesunden Sportler*innen ein schwerer Verlauf stattfinden und die Möglichkeiten, gerade für Mannschaftssport oder Training, das vornehmlich in Fitnesscentern stattfindet, wurden durch die Pandemieeinschränkungsmaßnahmen erheblich reduziert \cite{Niess2020}. Laddu et al. betonen, dass gerade sportliche Inaktivität den Verlauf einer möglichen Covid-Erkrankung erschweren kann \cite{Laddu2021} und Lippi et al. verschärfen diese Einschätzung als ``Kaleidoskop unvorteilhafter metabolischer Effekte, die das Risiko vieler schwerer Erkrankungen dramatisch erhöhen können'' \cite{Lippi2020}. %‘a kaleidoscope of unfavourable metabolic effects that would dramatically increase the risk of many severe and disabling disorders’
Auch hat der pandemiegeschuldete Rückgang sportlicher Aktivitäten deutliche Auswirkung auf die Psyche der Betroffenen \cite{Claussen2020, Brooks2020, Mutz2021b}.

Interventions- und/oder Gegenmaßnahmen für Sportler*innen aller Sparten scheinen daher dringend erforderlich.
Die World Health Organization (WHO) empfiehlt mehrere Maßnahmen zum Entgegenwirken der pandemiebedingten Rückgänge, unter anderem auch die Verwendung von Online-Videoangeboten oder -konferenzen \cite{WHO2020}.
Goecke betont die durch die Pandemie aufkommende Chance auf Aufschwung der Digitalisierung des Sports, insbesondere als Faktor der Suchtprävention \cite{Goecke2020}.
Im Bezug auf sport- und physiotherapeutische Behandlung während der Pandemie diskutiert Keller insbesondere E-Health, Telerehabilitation, Telemedizin oder Onlinebehandlung \cite{Keller2021}.
In der Sportlehre an Schulen und Hochschulen haben digitale Transformationsprozesse den Unterricht bereits bereichert oder erweitert \cite{Wendeborn2019, Roth2020, Mierau2020}. Als dafür hilfreich  genannt werden unter anderem Smartphone- oder Tabletapps und Videoanalysen bzw. -feedback \cite{Buchegger2016}.
Bentlage et al. fassen den Stand der Forschung im Frühstadium der Pandemie zusammen und nennen Empfehlungen und Programme der WHO, American Heart Association (AHA) und verwandten Arbeiten, wie Exergames, Maßnahmen aus dem Astronautentraining oder Tai Chi \cite{Bentlage2020}. 
Kauer et al. merken an, dass die Reaktionen auf die Pandemie bei einigen Altersgruppen sogar die Häufigkeit des Sports im Freien erhöhen kann und darüber hinaus digitale Anwendungen zuträglich für die Trainingshäufigkeit sein können \cite{KauerBerk2020}.

In diesem Zuge untersuchten Sportwissenschaftler*innen und Forscher*innen verwandter Disziplinen bereits den Einsatz digitaler Medien und/oder Trainingsanwendungen seit Beginn der Pandemie zwecks Aufrechterhaltung von Sport und Training.
Hafner und Busch intervenierten in die Einschränkungsmaßnahmen durch Online-Videokonferenzen von Hochschullehrkräften und Sportlehramtsstudierenden und stellten einen Rückgang sozialer Defizite fest, erwähnen aber auch das Risiko digitaler Sportalternativen für körperliche Mobilität und physisches sowie psychisches Befinden \cite{Hafner2021}.
Vereine in Deutschland haben bereits virtuelle Trainings und/oder Online-Angebote genutzt oder angeboten, was das lokale Angebot vieler Sportarten allerdings nicht ersetzen kann \cite{Kehl2021} -- Ehnold et al. bestätigen die Nutzung der Digitalisierung zur Kommunikation und Verwaltung, nicht aber zur Unterstützung des Sports an sich \cite{Ehnold2019}.
Vickey et al. und Martin et al. verglichen populäre Fitnessapps und präsentierten verschiedene Vorzüge für Sport und Fitness \cite{Vickey2013, Martin2015}, Depper und Howe unterstrichen die Vorteile von Apps und anderen digitalen Technologien zur Sportunterstützung, kritisierten im gleichen Zuge allerdings auch den Rückgang sozialer und interaktiver Elemente durch diese \cite{Depper2017}.
Tu et al. fügen hinzu, dass nicht allein Motivationsfeatures oder Gamifizierung die Langzeitnutzung von Sportapps begünstigen, sondern vor allem soziale Komponenten förderlich für kontinuierliches Trainingsverhalten sind \cite{Tu2019}.
Valcarce-Torrente et al. fanden allerdings keinerlei signifikante Unterschiede zwischen Fitnessappnutzer*innen und Nichtnutzer*innen in Bezug auf Sportverhalten, Zufriedenheit oder Fitnesscenternutzung \cite{Valcarce-Torrente2021}.
Yang und Koenigstorfer stellten einen signifikanten Rückgang sportlicher Aktivität bei Befragten aus den USA fest und verglichen demnach eine Kontrollgruppe zu Gruppen mit verschiedenen Ausprägungen von Fitnessappnutzung. Sportliche Aktivität nahm daher schwächer ab, wenn Teilnehmer*innen Fitnessapps benutzen, was mit dem Einsatz von Gamifizierung darüber hinaus noch verbessert wurde \cite{Yang2020}.
Corregidor-Sánchez et al. heben die Chancen von Exergames gerade im Zeitalter von Lockdowns und pandemischen Einflüssen hervor \cite{Corregidor2021}, Rüth und Kaspar betonen außerdem die Vorteile und Limitationen von pädagogischen und sozialen Exergames dieser Zeiten \cite{Rueth2021}.
Chtourou et al. diskutieren die Wichtigkeit sportlicher Aktivität während Lockdowns und Quarantäne, nicht allein wegen der Abträglichkeit persönlicher Fitness, sondern auch in Hinsicht auf Stress-, Angst- und/oder Depressionsgefahr, dem Heimtraining, Exergamess, Tanz oder Yoga entgegenwirken kann \cite{Chtourou2020}.
McDonough et al. berichten von einer Trainingsintervention via YouTube, die Muskelkraft und Trainingsverhalten messbar erhöhte \cite{McDonough2021}, im Setting von Grundschulsport konnten Taufik et al. ähnliche Ergebnisse bestätigen \cite{Taufik2021}.

Zusammengefasst eröffnen sich durch digitale Trainingsanwendung potentielle Chancen und Möglichkeiten, Sport und Training aufrechzuerhalten, pandemischen Einschränkungsmaßnahmen entgegenzuwirken und deren negative Auswirkungen auf die persönliche Gesundheit und Fitness zu mindern. Zum Zeitpunkt des Erstellens der in dieser Arbeit präsentierten Studie gab es allerdings -- nach bestem Wissen -- noch keine verwandten Arbeiten, die das Trainingsverhalten über den Zeitraum der Lockdown- und Nicht-Lockdownphasen (in Deutschland) erfasst und mit der Nutzung digitaler Trainingsanwendungen (vgl. Abschnitt \ref{sec:DTA}) in Verbindung gebracht hat. Daher wird das Sport- und Nutzungsverhalten digitaler Trainingsanwendungen von ($n=108$) Sportler*innen aus Deutschland im Kontrast effektiver und ausgesetzter Einschränkungsmaßnahmen erfasst. Tabelle \ref{tab:Perioden} fasst die kritischsten Zeiträume der Pandemie in Deutschland zusammen \cite{Imoehl2021}, einschließlich des ersten Lockdowns (P2), des ``Lockdown lights'' (P4), der ``Corona-Notbremse'' (P6) und allen Zwischenzeiträumen. Zwischen diesen Perioden gaben die Befragten Trainingshäufigkeit, -dauer, Fitnesscenternutzung und Nutzung digitaler Trainingsanwendungen (genauer: Fitnesstracker, Fitnessapps, Smartwatches, Videoangebote, Videokonferenzen und Exergames) an. Darüber hinaus erwähnten sie nötige Funktionen, was diese Trainingsanwendungen wichtig für die Durchführung ihres Sports macht oder welche benötigt werden, damit dieses effektiv unterstützen kann. Somit bietet diese Arbeit einen Eindruck über den Verlauf der Trainingsaktivität deutscher Sportler*innen, deren Nutzung und Einstellung zu digitalen Trainingsanwendungen und entdeckt Wünsche und Vorstellungen neuer Funktionen, die die Verbindung von Sport und Digitalisierung verstärken und sportliche Aktivitäten auch in Ausnahmezeiten fördern können.
%%%%%%%%%%%%%%%%%%%%%%%%%%%%%%%

\begin{table}[h]
\begin{center}
\caption{Lockdown-Perioden während der Covid-19-Pandemie}\label{tab:Perioden}%
\begin{tabular}{@{}l  l  l@{}}
%\toprule
\textbf{P1} & Vor 1. Lockdown & Bis März 2020 \\
\midrule
\textbf{P2} & 1. Lockdown & März - Mai 2020 \\
\midrule
\textbf{P3} & Zwischen 1. \& 2. Lockdown & Juni - November 2020 \\
\midrule
\textbf{P4} & 2. Lockdown & November 2020 - Februar 2021\\
\midrule
\textbf{P5} & Zwischen 2. \& 3. Lockdown & Februar - April 2021\\
\midrule
\textbf{P6} & 3. Lockdown & April - Mai 2021 \\
%\botrule
\end{tabular}
\end{center}
\end{table}

Schließlich versucht diese Arbeit, die folgenden Forschungsfragen mittels quantitativer und qualitativer Daten zu beantworten:

\begin{itemize}
    \item „Wie sehr hat sich das Trainingsverhalten von Sportler*innen während der Lockdowns und Zwischenzeiträume in Bezug auf Häufigkeit, Dauer und Fitnesscenternutzung verändert?“
    \item „Welche digitalen Trainingsanwendungen benutzen Sportler*innen zur Unterstützung ihres Trainings und wie sehr hat sich die Nutzung dieser über die jeweiligen Zeiträume verändert?“
    \item „Welche (bestehenden oder erwünschten) Funktionen/Aspekte digitaler Trainingsanwendungen geben Sportler*innen als wichtig an, um ihr Training aufrecht zu erhalten?“
\end{itemize}

Als zentrale, übergeordnete Forschungsfrage behandelt sie also:
\begin{itemize}
    \item „Welche Funktionen/Aspekte sollten digitale Trainingsanwendungen aufzeigen, um den Bedarf der Nutzer*innen auch in Episoden ohne primäre Trainingsmöglichkeit zu erfüllen, körperliches Fitnesstraining zu gewährleisten?“
\end{itemize}

\subsection{Digitale Trainingsanwendungen}
\label{sec:DTA}
Als digitale Trainingsanwendungen werden in diesem Kontext Soft- und/oder Hardwarelösungen verstanden, die Sport, Fitness, Training oder generelle Gesundheitsmaßnahmen vorgeben, aufzeichnen, unterstützen und/oder gamifizieren. Um die Nutzung verschiedener Ansätze unterscheiden zu können, wurden diese in einzelne (nicht notwendigerweise erschöpfende) Kategorien unterteilt, die im Folgenden näher abgegrenzt werden. 
\subsubsection*{Fitnesstracker}
\textit{Fitnesstracker} (auch: Activity Tracker) bezeichnen tragbare elektronische Geräte zur Aufzeichnung fitness- und gesundheitsrelevanter Daten wie Laufstrecken, Energieverbrauch, Pulsfrequenz oder Schlafdaten (z.B. FitBit, Xiaomi Band, Garmin vivofit). Explizit ausgeschlossen von dieser Kategorie wurden \textit{Smartwatches}, da diese meist technisch umfangreicher sind und gekoppelt werden können, sowie sich in Größe und Form deutlich unterscheiden, was einen Unterschied in der Verwendung ausmachen kann.
\subsubsection*{Fitnessapps}
\textit{Fitnessapps} sind Anwendungen, die auf mobile Geräte heruntergeladen und verwendet werden können, um fit zu werden oder zu bleiben. Hierbei sind vorrangig Apps gemeint, die das tatsächliche Training betreffen, aber auch generellere Apps wie z.B. Kalorienrechner können angegeben werden (z.B. Nike+ Training Club, Adidas Training, MyFitnessPal).
\subsubsection*{Smartwatches}
\textit{Smartwatches} bezeichnen elektronische Armbanduhren, die über zusätzliche Sensoren, Aktuatoren sowie Computerfunktionalitäten und -kopplungen verfügen können (z.B. Garmin Fenix, Apple Watch, Samsung Galaxy Watch). Darüber hinaus können Nutzer*innen zusätzliche Funktionen und Programme individuell hinzufügen oder konfigurieren.
\subsubsection*{Videoangebote}
\textit{Videoangebote} können Übungsdurchführung oder Trainingskompositionen demonstrieren oder anleiten. Hierbei beziehen wir uns auf nicht-interaktive Videos (z.B. spezialisierte YouTube-Kanäle).
\subsubsection*{Videokonferenzen}
\textit{Videokonferenzsysteme} ermöglichen Audio- und Videokommunikation über das Internet. Trainingssitzungen und Übungen können somit auch über Distanz geteilt werden (z.B. Zoom, Skype, FaceTime).
\subsubsection*{Exergames}
Schließlich verbinden \textit{Exergames} Übungen, Ausdauertraining oder andere sportliche Aktivitäten mit Videospielelementen. Diese können lose auf Bewegungserkennung basieren oder individuell auf konkrete Übungsdurchführung angepasst sein (z.B. WiiFit, Ring Fit Adventure, Kinect Sports Rivals).
Um eine möglichst scharfe Trennung dieser Kategorien zu ziehen, wurden den Teilnehmern diese während des Fragebogens mit Beispielgeräten oder -anwendungen, Bildern und Erläuterungstext vorgestellt.

\newpage
\section{Methodik}
Um die oben genannten Forschungsfragen beantworten zu können, wurde ein spezifischer Fragebogen entwickelt, der das Trainingsverhalten von Sportler*innen während der Covid-19-Pandemie erfasst, die Einstellung und Nutzung zu den verschiedenen Kategorien digitaler Trainingsanwendungen misst und Unterschiede hinsichtlich dieser Entwicklungen auch zwischen Hauptsporttypen (Fitness, Kraftsport, Mannschaftssport) feststellen kann. Dieser wurde von der Ethikkommission der Universität einiger der Autoren genehmigt und über einen Zeitraum von 4 Wochen (im Juni 2021) direkt nach dem Ende des letzten behandelten Lockdownabschnitts (vgl. Tabelle \ref{tab:Perioden}) veröffentlicht, um die Aktualität der Daten zu gewährleisten und wurde über Online-Umfrageportale und Email-Verteiler vertrieben. Die folgenden Abschnitte befassen sich mit der Beschreibung, Analyse und Interpretation dieses Fragebogens.

\subsection{Fragebogen}
Nach einer Einleitung über den Zweck, Inhalt und Ablauf der Studie sowie der Auflärung über Datenschutz, Kontaktmöglichkeiten und der Einverständniserklärung wurde der eigentliche Fragebogen präsentiert.
Dieser bestand aus einem Abschnitt über die generelle sportliche Aktivität der jeweiligen Hauptsportart, Fragen über die Häufigkeit (0-7 mal pro Woche), Dauer und Fitnesscenternutzung für jede Zeitperiode (siehe Tabelle \ref{tab:Perioden}) und Detailfragen über die Nutzung und Einstellung zu digitalen Trainingsanwendungen. Letztere umfassten Fitnesstracker, Fitness-Apps, Smartwatches, Videoangebote, Videokonferenzen sowie Exergames und erhoben bei bestätigter Nutzung die Häufigkeit (für jede Zeitperiode), Einfachheit der Bedienung und subjektive Nützlichkeit über 7-stufige Likert-Skalen. Darüber hinaus folgten qualitative Fragen, welche Funktionen dem Teilnehmer besonders wichtig sind oder welche (bislang noch nicht existierenden) Funktionen wünschenswert wären. Bei verneinter Nutzung beantwortete der Teilnehmer stattdessen, warum man sich dagegen entschieden hat, ob man sich vorstellen könnte, die digitale Trainingsanwendung in Zukunft zu nutzen und welche Funktionen dafür gewährleistet werden müssten. Abschließend erfasste der Fragebogen demografische Randdaten wie Alter, biologisches Geschlecht und den höchsten Bildungsabschluss und bot an, subjektive Anmerkungen zu Training während der Pandemie, dem Einsatz digitaler Trainingsanwendungen und generelle Kommentare hinzuzufügen.

\subsection{Teilnehmer}
Nach Ausschluss ungültiger Antworten nahmen ($n=108$) Teilnehmer*innen im Alter von $15-57$ Jahren ($M=28,1; SD=9,3$) an der Befragung teil ($53,7\%$ biologisch weiblich, $33\%$ männlich). $52,8\%$ von ihnen besaßen einen Hochschulabschluss, $15,7\%$ eine Ausbildung und $14,8\%$ gaben den Schulabschluss als höchsten Bildungsabschluss an.

\subsection{Kategorisierung}
Jeder der Teilnehmenden wurde aufgrund der eingangs genannten Hauptsportart in die Kategorie Fitness, Kraftsport, Mannschaftssport oder Andere eingeordnet, um spätere Schlüsse über Unterschiede zwischen Sporttypen ziehen zu können. Fitness beinhaltete Antworten wie Fitness, Ausdauer, Laufen und ähnliche Sportarten, Kraftsport bezeichnete Kraftsport, Krafttraining, Bodybuilding, etc. und Mannschaftssport umfasste Fußball, Handball, Basketball und verwandte Sportarten. Einzelnennungen oder Sportarten, die in dieser Kategorisierung nicht zugeordnet werden konnten, wurden unter Andere zusammengefasst (z.B. Reiten, Fechten, Bouldern), aufgrund eines zu unterschiedlichen Bilds allerdings nicht in die Gruppenvergleiche miteinbezogen.

\subsection{Analyse}
Um die einzelnen Gesichtspunkte der Forschungsfragen zu behandeln, wurden den Datenformaten entsprechende, ausschlaggebende statistische Verfahren herangezogen. Vergleiche zwischen einzelnen Perioden (hinsichtlich Häufigkeit, Nutzung oder Dauer) greifen auf (gepaarte) zweiseitige Studentsche $t$-Tests \cite{TTest} zurück, bei mehreren zeitlich aufeinanderfolgenden Paaren erst nach einer übergreifenden ANOVA \cite{ANOVA}. Für nicht-normalverteilte Datensätze wurde stattdessen ein Wilcoxon-Test herangezogen \cite{Wilcoxon}. Um die Bedeutsamkeit eines Unterschieds zu quantifizieren, wurde bei signifikanten Unterschieden die Effektstärke nach Cohens \textit{d} berechnet und interpretiert \cite{CohenD}. Qualitative Antworten wurden anhand strukturierender Inhaltsanalysen nach Mayring \cite{mayring2010qualitative} kategorisiert und kumuliert.

\section{Ergebnisse}

\begin{figure}[h]%
\centering
\includegraphics[width=0.9\textwidth]{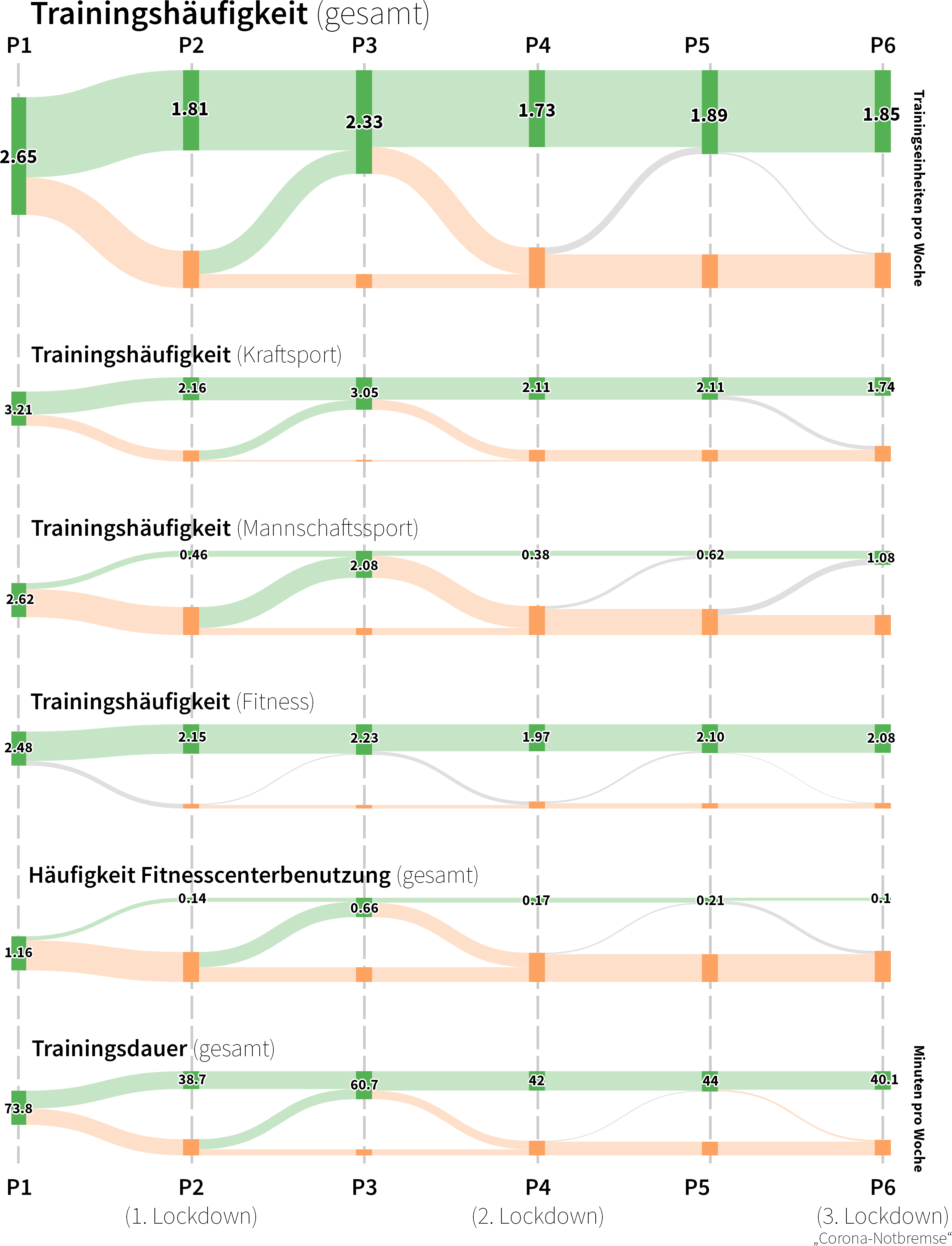}
\caption{Entwicklung der Trainingshäufigkeit (in Einheiten pro Woche) und Trainingsdauer (in Minuten pro Woche) während der Lockdownphasen der Covid-19-Pandemie. Grüne Übergänge bezeichnen einen signifikanten Anstieg, rote einen signifikanten Abfall ($p<0.05$) im Bezug auf die vorangegangene Periode, wobei die Breite jeweils proportional zur absoluten Häufigkeitsänderung ist.}~\label{fig:sankey}
\end{figure}

Die Häufigkeit der Trainingsaktivitäten hat mit Beginn des ersten Lockdowns signifikant abgenommen ($p<0.05, d=0.51$, siehe Abb. \ref{fig:sankey}).
Auf Sportgruppen heruntergebrochen zeigte sich dies bei sowohl Kraftsportler*innen und Mannschaftssportler*innen ($p<0.05, d_{Kraft}=0.59, d_{Mannschaft}=1.55$), allerdings nicht bei Sportler*innen, die vorrangig Fitness trainieren ($p>0.05$).
Nach der Aufhebung des ersten Lockdowns erhöhte sich diese Häufigkeit wieder signifikant ($p<0.05, d_{Gesamt}=0.33, d_{Kraft}=0.49, d_{Mannschaft}=1.22$), bis auf die Untergruppe der Fitnesssportler*innen ($p>0.05$).
Die Einführung des zweiten Lockdowns führte ebenso zur Abnahme der Häufigkeit ($p<0.05, d_{Gesamt}=0.38, d_{Kraft}=0.51, d_{Mannschaft}=1.38$) bis auf im Fitness ($p>0.05$).
Nach diesem änderten sich die Häufigkeiten allerdings nicht mehr ($p>0.05$), sondern blieben auf dem niedrigen Niveau des zweiten Lockdowns, sowohl vor als auch während des dritten Lockdowns.

\newpage
\newpage

Bezogen auf die Nutzung von Fitnesscentern nahm die Häufigkeit mit Einführung des ersten Lockdowns stark ab ($p<0.05, d=0.84$). 
Dies zeigte sich auch im Speziellen für Kraftsportler*innen und Fitnesssportler*innen ($p<0.05, d_{Kraft}=2.03, d_{Fitness}=0.79$), nicht aber für Mannschaftssportler*innen ($p>0.05$).
Gleichermaßen stiegen diese auch nach dem ersten Lockdown an ($p<0.05, d_{Gesamt}=0.51, d_{Kraft}=1.05, d_{Fitness}=0.4$) und nahmen zum Zweiten wieder ab ($p>0.05, d_{Gesamt}=0.45, d_{Kraft}=1.21$), wonach sie ebenso auf diesem Niveau blieben ($p>0.05$).

Die Dauer der wöchentlichen Trainings ist während des ersten Lockdowns für alle Sportlergruppen sowie gruppenübergreifend zurückgegangen ($p<0.05, d_{Gesamt}=0.85, d_{Kraft}=1, d_{Mannschaft}=1.51, d_{Fitness}=0.49$), danach signifikant gestiegen ($p<0.05, d_{Gesamt}=0.43, d_{Kraft}=0.39, d_{Mannschaft}=0.89, d_{Fitness}=0.15$) und zum Zweiten wieder gesunken ($p<0.05$), selbst leicht bei Fitnessportler*innen, deren Häufigkeit sich während der Pandemie nicht signifikant veränderte.
Zwischen dem zweiten und dritten Lockdown gab es keine signifikanten Veränderungen ($p>0.05$).
Letztlich wurde eine weitere leichte, signifikante Verringerung der Dauer durch den Dritten Lockdown gemessen ($p<0.05, d=0.09$), die allerdings nicht in einzelnen Gruppen reflektiert wurde ($p>0.05$).

\begin{figure}[h]%
\centering
\includegraphics[width=0.5\textwidth]{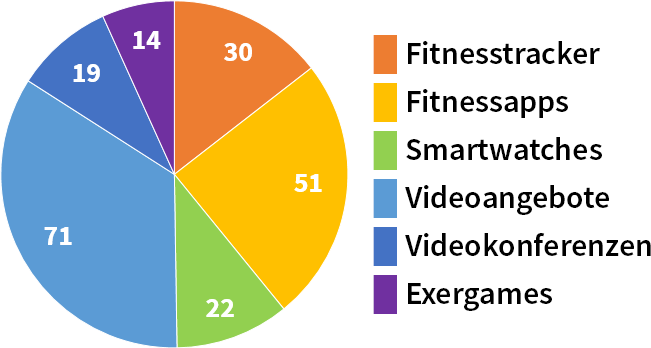}
\caption{Nutzung digitaler Trainingsanwendungen (absolut, Mehrfachnennungen möglich).}\label{fig:DTApie}
\end{figure}

Bezüglich digitaler Trainingsanwendungen waren Videoangebote unter allen Befragten am populärsten ($65,7\%$, vgl. Abb. \ref{fig:DTApie}). Danach folgten Fitnessapps ($47,2\%$), Fitnesstracker ($27,8\%$), Smartwatches ($20,4\%$), Videokonferenzen ($17,6\%$) und zuletzt Exergames ($13\%$).
Dieselbe Rangfolge zeigte sich bei Fitnesssportler*innen, von denen sogar $75,8\%$ Videoangebote nutzen.
Kraftsportler*innen hingegen bevorzugten Fitnessapps ($68,4\%$) vor Videoangeboten ($52,6\%$), Smartwatches ($42,1\%$), Fitnesstrackern ($36,8\%$) und Exergames ($15,8\%$). Keiner der befragten Kraftsportler*innen nutzte Videokonferenzangebote.
Unter den Mannschaftssportler*innen lagen Fitnesstracker, Fitnessapps und Videoangebote gleichauf ($46,2\%$). Smartwatches und Videokonferenzen wurden hier selten angegeben ($15,4\%$), Exergames nie.
\newpage

\begin{figure}[h]%
\centering
\includegraphics[width=0.9\textwidth]{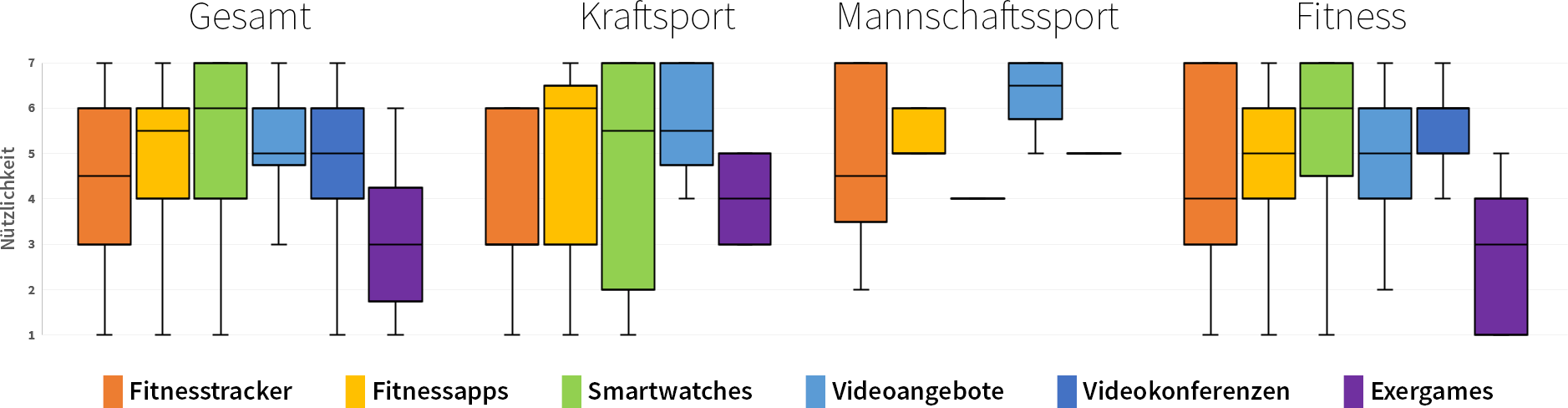}
\caption{Nützlichkeit digitaler Trainingsanwendungen nach Gruppen. Die Kastengrafik bildet Mediane (--), Quartile (Kasten) und Spannweite (Whisker) ab.}\label{fig:DTAHilfreichBoxplot}
\end{figure}

Videoangebote wurden darüber hinaus unter allen Teilnehmenden als am Hilfreichsten bewertet ($M=5,26$, vgl. Abb. \ref{fig:DTAHilfreichBoxplot}).
Danach folgten Smartwatches ($M=5,09$), Videokonferenzen ($M=5,00$), Fitnessapps ($M=4,98$), Fitnesstracker ($M=4,37$) und Exergames ($M=3,21$).
Kraftsportler*innen und Mannschaftssportler*innen teilten diese Einschätzung, Fitnesssportler*innen stuften allerdings Smartwatches ($M=5,56$) und Videokonferenzen ($M=5,36$) als hilfreicher ein.
Exergames wurden allerdings als am Einfachsten zu bedienen angegeben ($M=6,21$), noch vor Videoangeboten ($M=6,13$), Fitnesstrackern ($M=6,07$), Smartwatches ($M=5,91$), Fitnessapps ($M=5,69$) und Videokonferenzen ($M=5,50$).
Kraft- und Fitnesssportler*innen schätzten jedoch Videoangebote als am Einfachsten ein $(M=6,8 ; M=6,11$), wobei Mannschaftssportler*innen Smartwatches und Videokonferenzen ($M=7$) vorzogen.

\begin{figure}[h]%
\centering
\includegraphics[width=\textwidth]{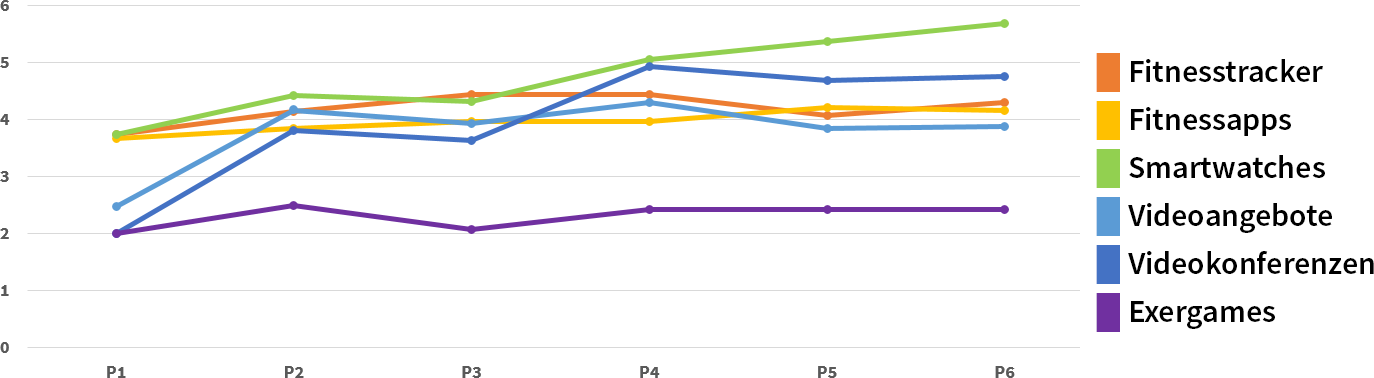}
\caption{Nutzungshäufigkeit von Besitzern digitaler Trainingsanwendungen über die Phasen der Covid-19-Pandemie.}\label{fig:DTAHaeufigkeitChart}
\end{figure}

Im Vergleich zu vor der Pandemie zeigten Sportler*innen signifikant höhere Nutzungen von Smartwatches, Videoangeboten und Videokonferenzen mit mittleren bis starken Effektgrößen ($p<0.05, d_{Smartwatches}=0.84, d_{Videoangebote}=0.63, d_{Videokonferenzen}=0.94$, vgl. Abb. \ref{fig:DTAHaeufigkeitChart}), während die Nutzung von Fitnessapps, Fitnesstrackern oder Exergames unverändert blieb ($p>0.05$).

\begin{figure}[h]%
\centering
\includegraphics[width=\textwidth]{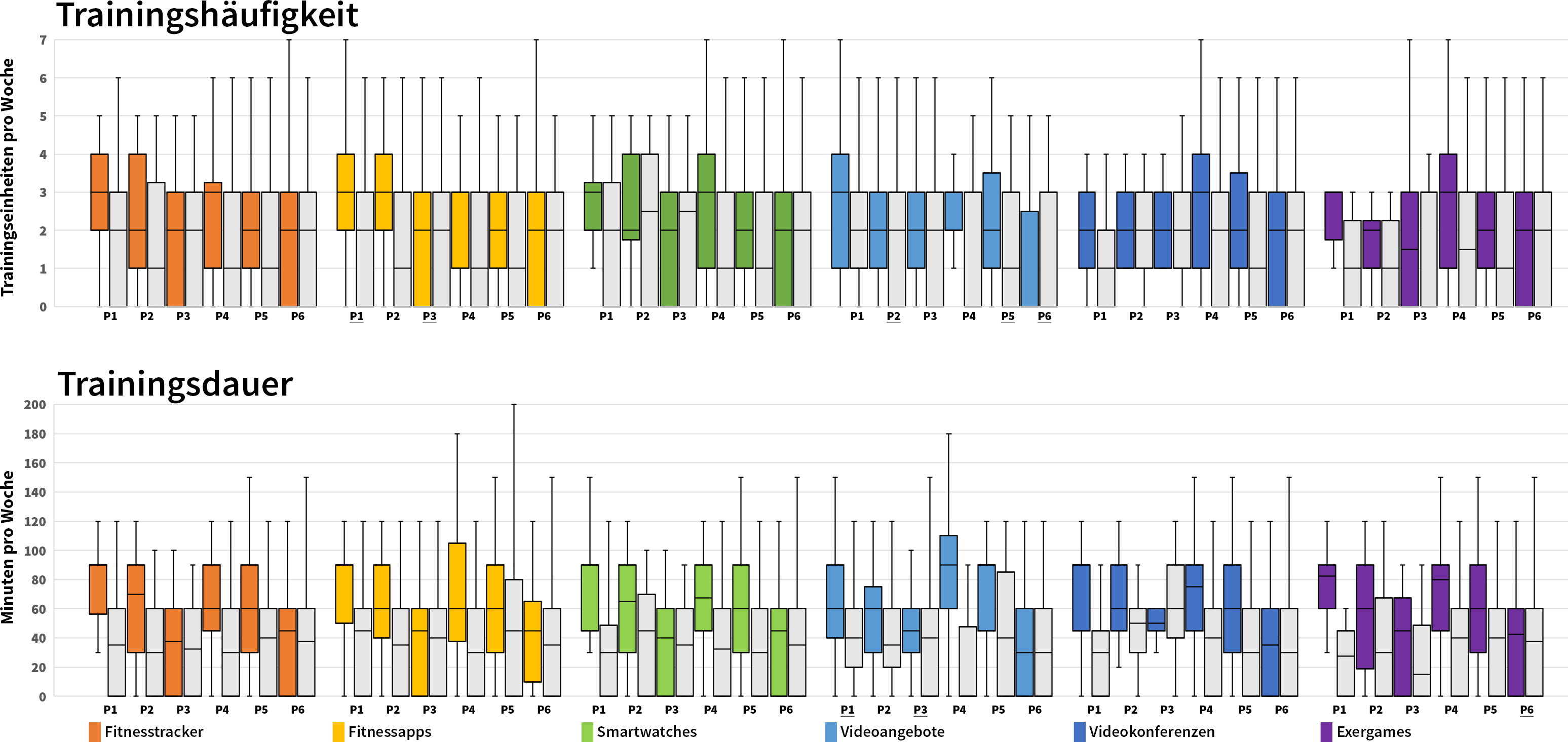}
\caption{Trainingshäufigkeit und -dauer von Teilnehmenden, die angaben, die jeweilige digitale Trainingsanwendung zu benutzen, im Vergleich zu Nichtnutzenden (grau), jeweils pro Periode. Die Kastengrafik bildet Mediane (--), Quartile (Kasten) und Spannweite (Whisker) ab.}\label{fig:MitVsOhne}
\end{figure}

\newpage

Im Vergleich von Nutzer*innen spezifischer digitaler Trainingsanwendungen zu denjenigen Teilnehmenden, die angaben, diese nicht zu verwenden, ergaben sich mitunter signifikante Unterschiede in Trainingshäufigkeit und -dauer (siehe Abb. \ref{fig:MitVsOhne}). So trainierten Nutzer*innen von Fitnessapps öfter in den Nicht-Lockdown-Phasen P1 und P3 ($p<0.05, d_{P1}=0.44, d_{P3}=0.45$), Nutzer von Videoangeboten allerdings häufiger in P2, P5 und P6 ($p<0.05, d_{P2}=0.43, d_{P5}=0.59, d_{P6}=0.48$), verglichen mit Nichtnutzenden. Außerdem dauerte das Training von Videoangebotsnutzern in P1 und P3 kürzer($p<0.05, d_{P1}=0.45,  d_{P3}=0.49$), das von Videokonferenznutzern in P6 allerdings länger ($p<0.05, d=0.67$).

\subsection{Qualitative}
\label{sec:Qualitative}
In Bezug auf die wichtigsten Funktionen von Fitnesstrackern gaben 13 Teilnehmer*innen die Pulsmessung an, 7 erwähnten den Schrittzähler, 7 GPS-Funktionen, 6 das Kalorienzählen, 5 generelle Aufzeichnung des Trainings, 3 Schlafdatenüberwachung, 2 die Stoppuhr und 1 Musikkontrolle. Gewünschte Zusatzfunktionen waren unter anderem Blutdruckmessung, Körperfettanteilmessung, Mannschaftsauswertung oder Musikwiedergabe passend zur Trainingsintensität. Teilnehmer*innen, die keine Fitnesstracker benutzen, sind generell nicht überzeugt (35), finden den Preis zu hoch (13), haben Datenschutzbedenken (10), bereits eine andere Alternative (8) oder keine Vereinbarkeit mit ihrem Sport (3). Nötige Funktionen, um zur Nutzung zu verleiten wären motivationssteigernde Maßnahmen (3), aufbereitete Gesundheitsinformationen, \textit{``Aufforderung[en] zur Pause in Abhängigkeit von Körperlicher Aktivität''} oder bislang nicht verfügbare Informationen, beispielsweise \textit{``Wie voll ist das Fitnessstudio?''}.\\
An Fitnessapps schätzen Benutzer*innen ebenfalls die Aufzeichnung der Kalorien (15) und des Trainings (14), allerdings auch Übungserläuterung (8). Des Weiteren wurden GPS-Tracking (7), Pulsmessung (6), Schrittzähler (3) und automatische Anpassung (2) genannt. Als Zusatzfunktionen wünschen sich Nutzer*innen quantitative Vergleiche zwischen Trainingseinheiten, Körperfettanteilmessung, Gegenwind-/Steigungsmessungen, die automatische Erkennung und Einschätzung von Kalorien anhand fotografierter Lebensmittel, Mannschaftsauswertung und Live-Kurse. Andere Teilnehmer*innen lehnten Fitnessapps generell ab (38), haben Datenschutzbedenken (5), lehnen die Nutzung wegen Kosten ab (1) oder dem Stress, den diese verursachen können (1). Als nötige Funktionen gaben sie an, Fitnessapps sollten \textit{``detaillierte Gesundheits- und Leistungsinformationen''} aggregieren, persönliche Beratung bieten und generell motivieren (3).\\
Nutzer*innen von Smartwatches verwenden diese hauptsächlich zur Pulsmessung (9) und Trainingsaufzeichnung (7), gelegentlich als Schlafdatentracker (2), Schrittzähler (1), GPS-Modul (1) oder zur Musikkontrolle (1). Darüber hinaus wünschen sie sich Blutdruck- und Blutzuckermessung, adaptive Musikwiedergabe passend zur Trainingsintensität und die Erkennung und Benachrichtigung, ab welchem Punkt im Training die Kalorienverbrennung beginnt. Der Großteil der Nichtnutzenden ist generell nicht überzeugt von Smartwatches (34), findet diese zu teuer (22), hat Datenschutzbedenken (7) oder fühlt sich durch die Uhr in ihrem Training behindert (6). Dennoch gaben sie an, dass motivationssteigernde Funktionen oder vollständige digitale Assistenten sie zur Nutzung überreden könnten.\\
Videoangebote werden fast ausschließlich zur Übungserläuterung beziehungsweise persönlichen Korrektur eingesetzt (21), nur 3 nannten explizit den Motivationsfaktor und 2 die begleitende Musik als Argument. Gerade hierzu bitten sie um flexible Musikeinstimmung oder voneinander getrennte Erklärungs- und Musikaudiokanäle zur Personalisierung (4). Andere wünschten sich ein übergreifendes System, das Trainingseinheiten unterschiedlicher Videoanbieter vereint, sich steigernde Module, geteilte Online-Sessions oder Korrekturen, bzw. \textit{``Benachrichtigungen bei falscher Ausführung, welche dem Körper schaden könnten''}. 16 Teilnehmer*innen gaben an, nicht überzeugt von Videoangeboten zu sein, 5 betonten die soziale Entfremdung, die mit der Digitalisierung des Sports einher geht, 3 fühlen sich durch die Bedienung der Videos in ihrem Training behindert und 3 greifen auf andere digitale Trainingsalternativen zurück. Als nötige Funktionen gaben Teilnehmer*innen an, dass individuelle Trainingsberatung Teil der Videoangebote sein müsste, dieses explizit Motivation anregen sollte und ein Teilnehmer wünschte sich eine automatische Stoppfunktion, sobald das Training pausiert werde.\\
In Videokonferenzen ist vorrangig Kollaboration (7) und die Möglichkeit auf Korrekturen (2) im Vordergrund. Nutzer*innen bitten darüber hinaus um Funktionen für synchronisiertes Videoabspielen, die Einbindung von Sensorik für detaillierteres Trainerfeedback, zentrale Musiksteuerung für alle Teilnehmer*innen oder \textit{``VR Funktionalität mit Ganzkörpertracking''}. 39 Teilnehmer*innen sind nicht davon überzeugt, dass Videokonferenzen ihr Training unterstützen können, 11 betonten das Unwohlsein durch die Benutzung und 5 gaben an, dass dies nicht mit ihrem Training vereinbar sei oder keine Angebote existieren.\\
Als Hauptgrund des Nutzens von Exergames gaben Teilnehmer*innen Spaß oder Motivation an (5), gefolgt von Übungserläuterung und Korrektur (3). Darüber hinaus wünschen sie sich haptisches Feedback, bessere Rückmelde-/Korrekturmechanismen oder die Integration und Kombination in und von anderen Exergames und digitalen Trainingsanwendungen. 29 Teilnehmer*innen lehnten die Nutzung von Exergames für ihr Training generell ab, 10 erwähnten die zu hohe Kostenschwelle, 9 mangelt es an Angeboten und 2 Teilnehmer*innen gaben an, dass auch Exergames zu sozialer Entfremdung führen können. Um Exergames für ihr Training einsetzbar zu machen, sollen diese in erster Linie die Motivation steigern (8) und präzise Korrekturen liefern (4), konkrete Funktionen wurden allerdings nicht genannt.

\section{Diskussion}
Um den Einfluss der Pandemiemaßnahmen auf das Trainingsverhalten von Sportler*innen zu messen, wurden Häufigkeit, Dauer und Fitnesscenternutzung jeweils pro Woche für jeden kritischen Zeitraum (siehe Tabelle \ref{tab:Perioden}) erhoben. Trivialerweise ließ die Fitnesscenternutzung in den ersten beiden Lockdowns drastisch nach, da diese in diesen Zeiträumen geschlossen hatten und höchstens Eigentümer*innen oder Personal Zugriff gehabt haben konnte. Allerdings blieb die Nutzung nach dem 2. Lockdown konstant, was sich mit der generellen Trainingshäufigkeit und -dauer deckt. Dies lässt erahnen, dass die einschneidenden Pandemiemaßnahmen die Einstellung zum Sport und den alltäglichen Tagesablauf derart beeinflusst haben, dass persönliches Training deutlich negativ betroffen wurde, was Teilnehmer*innen in qualitativen Kommentaren bekräftigen. Den Mannschaftssport trafen die Kontakteinschränkungen in den Lockdowns am härtesten, aber auch Kraftsportler*innen wurden durch den verwehrten Fitnesscenterzugang deutlich eingeschränkt. Fitnesssportler*innen konnten sich am besten an die Schließung der Fitnesscenter anpassen, da sie ihr Training umgestalteten, sodass sie es allein zuhause oder im Freien durchführen konnten. Diese Veränderung in den Trainingsabläufen führte allerdings auch bei Fitnesssportler*innen zu einer signifikanten Verringerung der Dauer. Zusammenfassend lässt sich die zugehörige Forschungsfrage folgendermaßen beantworten:
\begin{itemize}
    \item Durch die Einführung der Pandemiemaßnahmen ist das Trainingsverhalten von Sportler*innen während der Lockdowns hinsichtlich Häufigkeit, Dauer und Fitnesscenternutzung signifikant zurückgegangen. Nach dem ersten Lockdown verbesserte sich das Trainingsverhalten entsprechend wieder, nach dem Zweiten blieb es allerdings auf einem konstant niedrigen Niveau.
\end{itemize}
\newpage

Im Bezug auf digitale Trainingsanwendungen, die dieser Entwicklung entgegenwirken oder neue Trainingsmöglichkeiten offenbaren könnten, haben sich Videoangebote und Fitnessapps am meisten hervorgetan. Dies ist reflektiert in dem hohen Anteil der Nutzer*innen, der subjektiven Nützlichkeit und Einfachheit der Bedienung. Dieser Vorsprung kann sich auch durch den einfachen Einstieg in diese Anwendungen erklären, da heutzutage beinahe jeder über Smartphones und/oder Computer verfügt, die sowohl Videoangebote abspielen als auch Fitnessapps ausführen können -- anders als Fitnesstracker, Smartwatches oder Exergames, die meist zusätzliche Hardwareanschaffungen benötigen. Auch wenn Videokonferenzen in der Pandemie für viele Anwendungsfälle populär und hilfreich geworden sind, lehnten Sportler*innen diese eher ab, da diese soziale Unbehaglichkeit hervorrufen können oder nicht mit dem Training vereinbar sind. Trotzdem stieg die Nutzung von Videokonferenzen, wie auch von Videoangeboten und Smartwatches im Vergleich zu vor der Pandemie deutlich an. Am wenigsten überzeugt waren die Befragten von Exergames, die weder oft genutzt, noch als nützlich für das eigene Training bewertet wurden. In einigen Lockdownphasen wurde höhere Trainingshäufigkeit (P2, P6 für Videoangebotsnutzer*innen) oder -dauer (P6 für Videokonferenznutzer*innen) im Vergleich zu Nichtnutzer*innen dieser Technologie gemessen. Zusammen mit dem Anstieg der Nutzung dieser Anwendungen lässt dies vermuten, dass digitale Trainingsanwendungen die Trainingshäufigkeit und -dauer der Nutzer*innen erhöhen können. Dementsprechend lässt sich die Frage nach digitalen Trainingsanwendungen wiefolgt beantworten:
\begin{itemize}
    \item Zur Unterstützung ihres Trainings benutzen Sportler*innen hauptsächlich Videoangebote und Fitnessapps. Bei Nutzer*innen von Videoangeboten, Smartwatches und Videokonferenzen konnte allerdings ein signifikanter Anstieg über den Pandemiezeitraum gemessen werden, während dieser keinen Einfluss auf die Nutzung anderer digitaler Trainingsanwendungen hatte. Die Nutzung von Videoangeboten oder -konferenzen konnte dazu beitragen, die Trainingshäufigkeit bzw. -dauer während Lockdownphasen zu erhöhen.
\end{itemize}

Generell gaben die Teilnehmer*innen an, dass das Tracking ihres Trainings und Körpers (Puls, Schritte, Kalorien, Schlaf, GPS), Übungserläuterung und -korrekturmaßnahmen sowie Motivation die wichtigsten bestehenden Funktionen unter allen Trainingsanwendungen seien. Die Möglichkeit der Kollaboration mit anderen Sportler*innen wurde auch erwähnt, fand bislang allerdings nur bei Videokonferenzen Anwendung. Zu ausnahmslos allen Trainingsanwendungen wurde angegeben, dass man sich weitere oder verbesserte motivationssteigernde Funktionen wünsche. Darüber hinaus wurden jeweils einige spezifische Vorschläge genannt, wie die Nützlichkeit von Trainingsanwendungen verbessert werden könnte oder die Nutzung für einige Sportarten und Sportler*innen überhaupt erst möglich gemacht würde (vgl. Abschnitt \ref{sec:Qualitative}). Die Auseinandersetzung der technischen Umsetzbarkeit und Relevanz dieser Vorschläge wird hier nicht im Detail erarbeitet, allerdings im Abschnitt \ref{sec:Ausblick} behandelt. 
Der Forschungsfrage entsprechend zusammengefasst:

\begin{itemize}
    \item Vor allem Körper- und Trainingsaufzeichnungsfunktionen, aber auch Übungserläuterung und Korrekturen stellen die wesentlichen Funktionen digitaler Trainingsanwendungen dar. Damit diese darüber hinaus den Sport oder Sportler*in unterstützen, sollten diese explizit Motivation fördern und Kollaboration ermöglichen.
\end{itemize}

Abschließend lassen sich die Erkenntnisse der bisherigen Angriffspunkte unter der übergreifenden allgemeinen Forschungsfrage zusammenfassen:

\begin{itemize}
    \item Die Nutzung einiger digitaler Trainingsanwendungen über den Zeitraum der Covid-19-Pandemie hat signifikant zugenommen, während die Trainingshäufigkeit, -dauer und Fitnesscenternutzung zurückging. Befragte bestätigten die Vorteile digitaler Trainingsanwendungen, um dem Rückgang an Trainingsmöglichkeiten entgegenzuwirken und griffen zumeist auf Angebote mit niedriger Einstiegsschwelle zurück (Videoangebote, Fitnessapps), die keine zusätzlichen Anschaffungen benötigten. Tracking, Übungserläuterung und Korrekturen können dabei helfen, fehlendes persönliches Feedback teilweise zu kompensieren und vor allem motivationssteigernde und Kollaborationsfeatures sind gefragt, die der verringerten sozialen Komponente des Trainings entgegenwirken können.
\end{itemize}
% Datenschutz
% In Verbindung mit related work?

\subsection{Limitationen und Ausblick}
\label{sec:Ausblick}
Die verhältnismäßig hohe Teilnahme von Akademiker*innen ($52,8\%$) an der Befragung kann zum Teil durch die Akquirierung der Teilnehmer*innen unter anderem über Online-Umfrageplattformen zustande gekommen sein, was die Heterogenität der befragten Population beeinflusst und bei der Interpretation der Ergebnisse berücksichtigt werden sollte.
Auch sollte an dieser Stelle noch einmal erwähnt werden, dass diese Umfrage nur Aussagen über in Deutschland lebende Sportler*innen, deren Trainingsverhalten über den Pandemiezeitraum und deren Vorliebe und Anforderungen an digitale Trainingsanwendungen treffen kann. Für eine detailliertere Analyse wurden diese Teilnehmer*innen in Untergruppen (Fitness, Kraftsport, Mannschaftssport) unterteilt, da sich diese als größte Überbegriffe herausstellten. Sportler*innen, die nicht in diese Kategorien passten, wurden von Gruppenvergleichen aufgrund von zu unterschiedlichen Sportarten ausgeschlossen, in der Gesamtanalyse allerdings berücksichtigt.
Der Vergleich von Nutzer*innen gegenüber Nichtnutzer*innen spezifischer Trainingsanwendungen und die Auswirkungen auf das Trainingsverhalten (vgl. Abb. \ref{fig:MitVsOhne}) kann lediglich Zusammenhänge darstellen, nicht aber unvoreingenommen vorhersagen, ob digitale Trainingsanwendungen die Trainingsintensität tatsächlich erhöhen. Eine längerfristige Studie, in der vorher nicht mit digitalen Trainingsanwendungen in Berührung gekommene Sportler*innen mittels Kontroll- und Interventionsgruppen beobachtet werden, hätte den Einfluss dieser Anwendungen genauer messen können, allerdings wurde aufgrund der besonderen Bedeutung der Zeitperiode eine retrospektive Befragung präferiert.

Im Weiteren nannten Teilnehmer*innen viele spezifische Features, die vertretbarerweise förderlich für Trainingsabläufe und technisch realistisch umsetzbar scheinen, gerade mit der verfügbaren Sensorik und Aktuatorik verbreiteter Smartphones. So könnten etwa \textit{``Aufforderung[en] zur Pause in Abhängigkeit von Körperlicher Aktivität''} durch einfache Videoerkennungsprotokolle gekoppelt mit (adaptiven) Trainingsprogrammen stattfinden, \textit{``Wie voll ist das Fitnessstudio?''} durch statistische Näherungen an übliche Besucherzahlen der Einrichtung (z.B. über Google Places API) beantwortet werden oder Gesundheitsübersichten durch Aggregation von Körper- und Trainingstrackingdaten angenähert werden. Auch \textit{``adaptive Musikwiedergabe passend zur Trainingsintensität''} ließe sich bereits mit einfacher Video- oder Körperdatensensorik abgleichen und gegebenenfalls sogar prozedural generierte Musik passend zu z.B. Puls oder Schrittgeschwindigkeit erzeugt und abgespielt werden. Aus Livedaten der Herzfrequenz lässt sich darüber hinaus auch die aerobe Schwelle schätzen und aktuelles Feedback über momentane Fettverbrennung präsentieren. Für komplexere Fragestellungen wie \textit{“Benachrichtigungen bei falscher Ausführung, welche dem Körper schaden könnten”} wären umfangreichere Übungsdatenbanken und/oder digitale physiologische Architekturen zur Erkennung und Korrektur notwendig, was allerdings im gewissen Rahmen auch über Skeletterkennung der Smartphonekamera oder Bewegungsmusteraufzeichnung über üblich verbaute Lage- und Beschleunigungssensoren funktioniert. Zu den vielmals erwähnten Motivationsfunktionen könnten aktuelle Ergebnisse aus dem Bereich der Gamifizierung und/oder der Exergames auf den nicht-spielerischen Kontext übertragen werden.

Zusammengefasst würden viele der notwendigen und gewünschten Funktionen in einer zentralen Smartphone-Fitnessapp untergebracht werden können. Diese sollte allerdings so generisch konzeptioniert sein, dass sie Videoangebote möglichst aller Quellen mit einschließt (da diese die populärste der erwähnten digitalen Trainingsanwendungen ist), Kollaborationsmechanismen wie Videokommunikation oder persönliche Betreuung anbietet, aber nicht voraussetzt, motivationsfördernde Anreize bietet und sowohl über das Smartphone (Kamera oder Sensorik) Echtzeitdaten aufzeichnen, aggregieren und interpretieren kann, also auch Kopplungen zu anderen Trackern (oder Smartwatches) ermöglicht, um weitere Körperdaten in die Analyse einfließen zu lassen.
% Soziale Komponente

\newpage
\section{Fazit}
Die SARS-CoV-2 Pandemie hat Leben und Alltag nahezu aller Menschen verändert. Dies zeigt sich auch in sportlichen Aktivitäten, die in vielen Ländern, Altersschichten und Sporttypen signifikant abgenommen hat. In Bezug auf deutsche Sportler*innen konnte diese Arbeit nach anderthalb Jahren und drei Lockdowns den Rückgang der Intensität bezüglich Häufigkeit und Dauer durch die Maßnahmen bestätigen und lässt einen konstanten langfristigen Rückgang erahnen, der sich nicht nur auf die Lockdowns beschränkt. Nichtsdestotrotz hat die Nutzung digitaler Trainingsanwendungen in diesem Zeitraum zugenommen, Sportler*innen öffnen sich mehr für neue Formen oder Erweiterungen des Trainings und geben in diesem Zuge ihre Vorstellungen und Wünsche an, wie diese in Zukunft noch besser unterstützen können. %Diese Arbeit lieferte Einsicht in das Trainingsverhalten über den bisherigen Pandemiezeitraum und 

% \section*{Interessenkonflikt}
% Der korrespondierende Autor gibt für sich und seine Koautoren an, dass kein Interessenkonflikt besteht.

% \begin{appendices}
% \section{Fragebogen} %% ?
% \end{appendices}

%\bibliography{sn-bibliography}
\end{document}